\documentclass[pdflatex,sn-mathphys-num]{sn-jnl}% Math and Physical Sciences Numbered Reference Style
\usepackage{graphicx}%
\usepackage{multirow}%
\usepackage{amsmath,amssymb,amsfonts}%
\usepackage{amsthm}%
\usepackage{mathrsfs}%
\usepackage[title]{appendix}%
\usepackage{xcolor}%
\usepackage{textcomp}%
\usepackage{manyfoot}%
\usepackage{booktabs}%
\usepackage{algorithm}%
\usepackage{algorithmicx}%
\usepackage{algpseudocode}%
\usepackage{listings}%
\theoremstyle{thmstyleone}%
\theoremstyle{thmstyletwo}%

\theoremstyle{thmstylethree}%

\begin{document}
	
	\title[Article Title]{Isochronous islands in the two-harmonic standard map}
	
	
	\author*[1]{\fnm{Michele} \sur{Mugnaine}}\email{mmugnaine@gmail.com}
	
	\author[1]{\fnm{Bruno B.} \sur{Leal}}\email{bruno.borges.leal@usp.br}
	\equalcont{These authors contributed equally to this work.}
	
	\author[2]{\fnm{Alfredo M.} \sur{ Ozorio de Almeida}}\email{alfredozorio@gmail.com}
	\equalcont{These authors contributed equally to this work.}
	
	\author[3,4]{\fnm{Ricardo L.} \sur{Viana}}\email{rlv640@gmail.com}
	\equalcont{These authors contributed equally to this work.}
	
	\author[1]{\fnm{Iber\^e L.} \sur{Caldas}}\email{ibere@if.usp.br}
	\equalcont{These authors contributed equally to this work.}

	\affil[1]{\orgdiv{Institute of Physics, University of S\~ao Paulo}, \orgaddress{\city{S\~ao Paulo}, \postcode{05508090}, \state{S\~ao Paulo}, \country{Brazil}}}
	
	\affil[2]{\orgdiv{Brazilian Center for Research in Physics}, \orgaddress{\city{Rio de Janeiro}, \postcode{22290-180}, \state{Rio de Janeiro}, \country{Brazil}}}
	
	\affil[3]{\orgdiv{Department of Physics}, \orgname{Federal University of Paran\'a}, \orgaddress{\city{Curitiba}, \postcode{82590-300}, \state{Paran\'a}, \country{Brazil}}}
	
	\affil[4]{\orgdiv{Interdisciplinary Center for Science, Technology, and Innovation}, \orgaddress{\city{Curitiba}, \postcode{81530-000}, \state{Paran\'a}, \country{Brazil}}}
	
	
	\abstract{Isochronous islands are regular solutions related to different chains of elliptic points but with the same winding number. These isochronous islands emerge in phase space as a response to multiple resonant perturbations and can be simulated using a simple discrete model called the two-harmonic standard map. We observed three types of isochronous transitions, which can be formed through saddle-node and pitchfork bifurcations.}

	
	
	
	\maketitle
	
	\section{Introduction}\label{sec1}
	
	In dynamical systems, isochronous islands are distinct regular solutions with same frequency which circulate distinct elliptic points with same period. In the plasma physics literature, such islands are called heteroclinic and recent studies present theoretical and experimental evidence of the emergence of such islands within magnetically confined plasmas in tokamaks \cite{bardoczi2021,evans2021}. In the plasma, for example, isochronous islands occur due to the interaction between multiple tearing modes which grow in the same rational surface, which lead to heteroclinic/isochronous bifurcations responsible for the emergence of the islands \cite{bardoczi2021}. The presence of isochronous islands are identified in several dynamical models related to nonlinear oscillators \cite{walker1969},  electron beam interactions with electrostatic waves\cite{deSousa2013},  periodic lattices\cite{lazarotto2022}, molecular physics \cite{carvalho1992} and, as already mentioned, plasma physics \cite{bardoczi2021,evans2021,leal2024,fraile2024}.

	These isochronous bifurcations can be simulated by a simple discrete model, named the two-harmonic standard map (THSM), proposed and analyzed in Ref. \cite{mugnaine2024} as a model for the competition between different resonant modes.  According to the Poincaré-Birkhoff theorem, any resonance with a rational winding number $r/s$ leads to the emergence of $2k$ periodic orbits with period $s$, with $k\in\mathbb{N}$,\cite{lichtenberg} which moves $r$ steps in the positive direction \cite{katok2003}. In this scenario, half of these orbits are unstable (hyperbolic points) and the other half is stable. If $k>1$, we have distinct islands with the same winding number, \textit{i.e.}, isochronous islands.
	
	It was shown that isochronous islands emerge in the phase space as a response to multiple resonant perturbations and the number of islands depends on the system's characteristic and the amplitude perturbations. The isochronous islands occurs in the same winding number surfaces, \textit{i.e.}, only modes with the same ratio as winding number can interact and cause the emergence of isochronous islands. The winding number $(r_1,s_1)$ can be different than $(r_2,s_2)$, but the ratios $r_1/s_1$ and $r_2/s_2$ must be equal \cite{leal2024}.  Besides the THSM, more complex models can also present the emergence of isochronous islands\cite{deSousa2013,leal2024}. As shown for the THSM, sadlle-node and pitchfork bifurcations formed the transitions responsible for the emergence of isochronous islands in the same frequency surface \cite{mugnaine2024}.
	
	In this paper, we consider and discuss isochronous bifurcations for other modes not presented in Ref. \cite{mugnaine2024}. In Section \ref{section:model}, we present the two-harmonic standard map model and discuss the role of each competing model in the system. The competition between the modes and the isochronous transitions are analyzed in Section \ref{section:transitions}. Our conclusions are presented in Section \ref{section:conclusions}.

	\section{The model}
	\label{section:model}
\begin{figure}[!h]
		\begin{center}
			\includegraphics[width=0.6\textwidth]{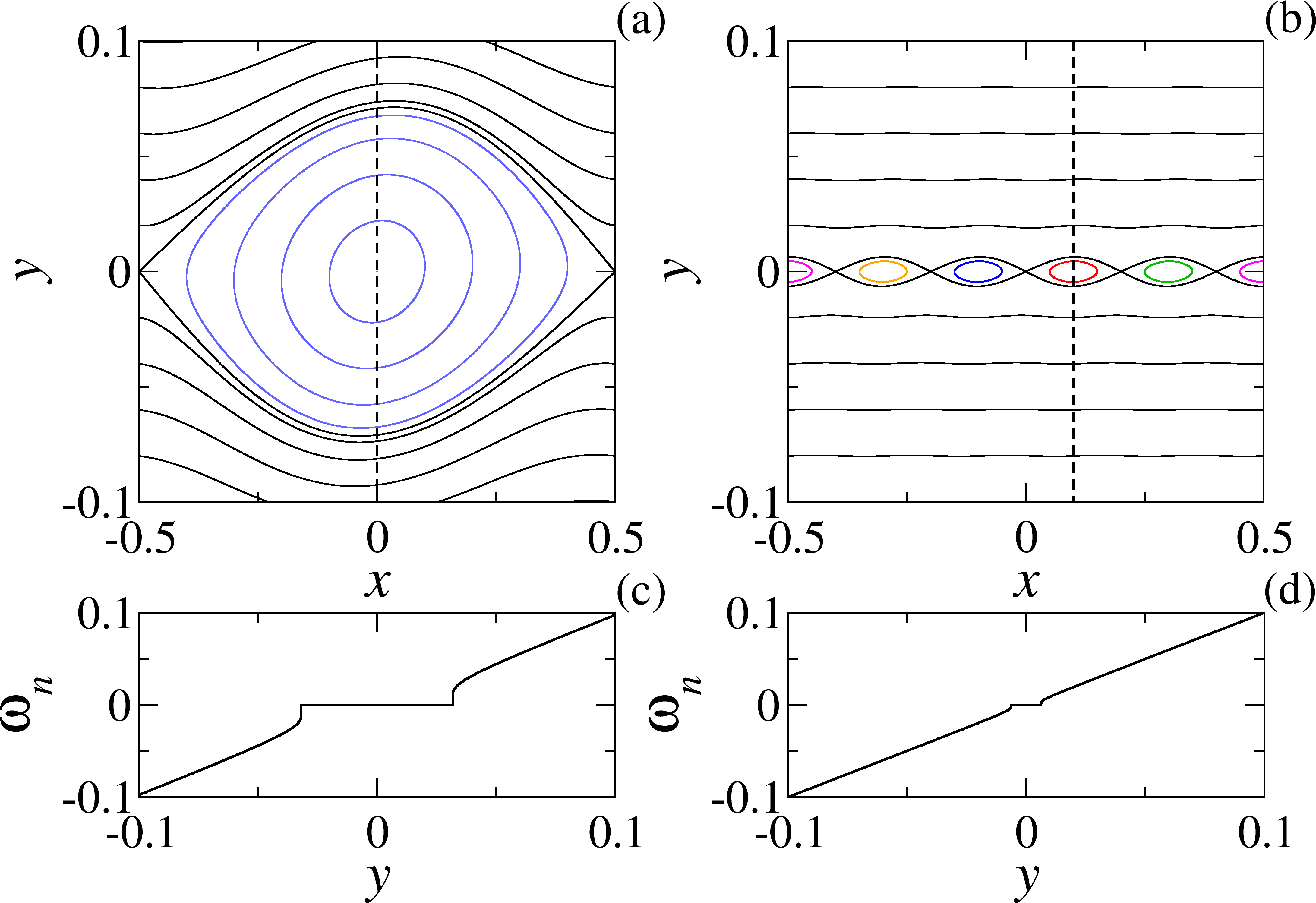}
			\caption{Phase spaces and winding number profile for isolated modes. In (a), we observe the phase space for $m_1=1$, $K_1=0.01$ and $K_2=0.0$. The winding number is computed in the dashed line and it is presented in panel (c). For (b), we have the isolated mode $m_2=5$, with $K_1=0$ e $K_2=0.01$. The respective winding number profile is shown in panel (d).}.
			\label{fig1}
		\end{center}
	\end{figure}

	The two-harmonic standard map is proposed as a generalization of the extended standard map, investigated in Refs. \cite{ketoja1989,satija1996,lomeli2006,greene1990,greene1987,ketoja1990,johannesson1988,black1990}, and it is defined by the equations\cite{mugnaine2024},
	\begin{eqnarray}
		\begin{aligned}
			x_{n+1}&=x_n+y_{n+1},\\
			y_{n+1}&=y_n-\dfrac{K_1}{2\pi m_1}\sin(2 \pi m_1 x_n) - \dfrac{K_2}{2 \pi m_2} \sin(2 \pi m_2 x_n),
		\end{aligned}
		\label{eq:ESNM}
	\end{eqnarray}
	where $K_1, K_2 \in \mathbb{R}$ and $m_1, m_2 \in \mathbb{N}$. The numbers $m_1$ and $m_2$ identify the modes of the system and depending on the amplitudes $K_1$ and $K_2$, the system can exhibit $m_1$ to $m_2$ islands, with $m_2 > m_1$. We take $\mod$ 1 for both variables in (\ref{eq:ESNM}). In order to understand the effect of each mode, represented by each sine function in (\ref{eq:ESNM}), we study each one separately by computing the phase spaces. Choosing $m_1=1$ and $m_2=5$ we have the phase spaces for each mode shown in Figure \ref{fig1}
	
	In Figure \ref{fig1} (a), we have the phase space for $m_1=1$, $K_1=0.01$ and $K_2=0$, \textit{i.e.}, we have just mode $m_1$. We observe an island of period 1 centered in $(x,y)=(0,0)$. For the mode $m_2$, we observe the phase space in Figure \ref{fig1} (b), with $K_1=0$ e $K_1=0.01$. In this case, we observe five distinct islands of period 1 in $y=0$. These islands are called isochronous, they have the same period but are distinct islands.
	
	With the phase spaces in Figure \ref{fig1} (a) and (b), we identify the effect of each mode in the system: the mode $m_{1,2}$ indicates the number of isochronous islands of period 1 in $y=0$. Along with the phase space, we also compute the winding number profile. The winding number is defined by the limit
	\begin{eqnarray}
		\omega_n=\lim_{n\to\infty} \dfrac{x_n-x_0}{n},
		\label{eq:wn}
	\end{eqnarray}
	where the variable $x_n$ is lifted to the real line. If the limit in (\ref{eq:wn}) is defined, we have a regular orbit, which can be periodic or quasi periodic. If the solution is chaotic, the limit does not converge. We compute the winding number for $10^4$ points at the lines depicted in Figure \ref{fig1} (a) and (b), \textit{i.e,} $x=0$ for panel (a) and $x=0.1$ for panel (b). The respective winding number profiles are shown in Figure \ref{fig1} (c) and (d), respectively.
	
	From the winding number profiles shown in Figure \ref{fig1} (c) and (d), we observe a monotonic increasing profile with a plateau in the island region. The winding number value $\omega_n=0$ is constant for the only island in Figure \ref{fig1} (a) and for the five distinct islands in Figure \ref{fig1} (b). With this, we say the islands are in the same winding number surface. As shown in Refs. \cite{mugnaine2024,leal2024,fraile2024}, it is necessary for islands with the same winding number to lie on the same rational surface for isochronous bifurcations to occur. As discussed with more details in Ref. \cite{mugnaine2024}, when  $K_1$ and $K_2$ are nonzero, we can have isochronous bifurcations, leading to transitions from mode $m_1$ to mode $m_2$. Now, we investigate the scenario with two modes acting on the systems, for nonzero $K_1$ and $K_2$.
	
	\section{Isochronous transitions}
	\label{section:transitions}
	A $m_1\to m_2$ transition occurs by isochronous bifurcations, \textit{i.e.}, the emergence of elliptic points on the same winding number surface, leading to the emergence of isochronous islands. With this in mind, we investigate the bifurcation diagrams for the fixed points in the phase space. Setting $K_1=0.05$, we find the fixed points of the systems for increasing values of $K_2$. We compute the bifurcation diagrams and phase spaces for the stages before and after the isochronous bifurcations. 
	
\begin{figure}[!h]
		\begin{center}
			\includegraphics[width=1.0\textwidth]{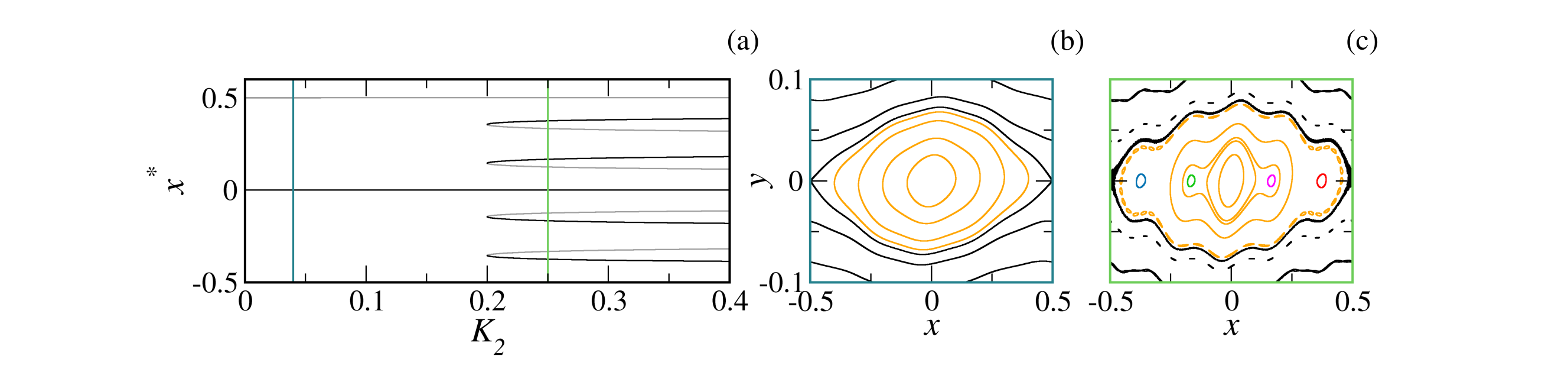}
			\caption{Transition from mode $m_1=1$ to $m_2=5$, with $K_1=0.05$. (a) The bifurcation diagram shows the evolution of the fixed points, where the elliptic and hyperbolic points are represented by black and gray points, respectively. We chose $K_2=0.04$ for the scenario before the isochronous bifurcations and $K_2=0.25$, for the scenario after it. Both values are highlighted by color vertical lines in panel (a). The respective phase spaces are show in panels (b) and (c), where the frame color is related to the color line in (a).}
			\label{fig2}
		\end{center}
	\end{figure}

	We initiate our analysis by the case of Figure \ref{fig1}, $m_1=1$ and $m_2=5$. The respective results are shown in Figure \ref{fig2}.
	
	In Figure \ref{fig2} (a), we have the bifurcation diagram for the fixed points of the case $m_1=1$ and $m_2=5$, with $K_1=0.05$. Up to $K_2=0.2$, we observe only one elliptic point and one hyperbolic points, the black and gray points at $x^*=0$ and $x^*=0.5$, respectively. At $K_2=0.2$ we have four saddle-node bifurcations and, then, four new elliptic points, as four hyperbolic points, emerge in the system.  All four bifurcations occur at the same value of $K_2$, leading to a transition with no intermediate modes.
	
	The phase space for the system before the transition is shown in Figure \ref{fig2} (b), for $K_2=0.04$. We observe only one island centered in the fixed point $(0,0)$. After the $1 \to 5$ transition occurred, we have the scenario depicted in Figure \ref{fig2} (c), where we observe five isochronous islands centered in elliptic points on line $y=0$. As show in the diagram, all fixed points emerge by saddle-node bifurcations, characterizing the first observed type of transition from mode $m_1$ to $m_2$.
	
\begin{figure}[!h]
		\begin{center}
			\includegraphics[width=1\textwidth]{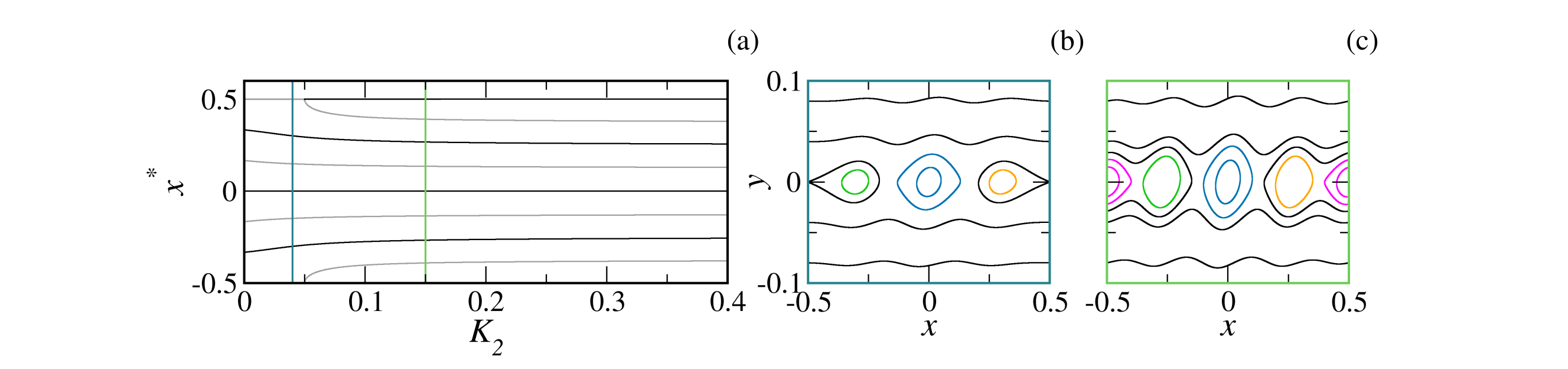}
			\caption{The $3 \to 4$ transition  by pitchfork bifurcation. In the bifurcation diagram in panel (a), we observe the occurrence of a pitchfork bifurcation ate $K_2=K_1=0.05$ where the hyperbolic point at $x^*=0.5$ becomes elliptic and two new hyperbolic point emerge. The scenario before the transition is shown in (b), with $K_2=0.04$, while the scenario after the transition is depicted in (c), with $K_2=0.15$.}
			\label{fig3}
		\end{center}
	\end{figure}

	Next, we analyze the case where $m_1=3$ and $m_2=4$. From the bifurcation diagram, shown in Figure \ref{fig3} (a), where we observe that, for $K_2<0.5$, there are only three elliptic points, indicating the predominance of mode $m_1=3$. For greater values of $K_2$, we have the emergence of two new elliptic points and the total of five elliptic points on the line $y=0$. We observe the occurrence of one pitchfork bifurcation, where the hyperbolic fixed point at $(0.5,0)$ becomes elliptic at the bifurcation and two new hyperbolic point emerge in the system.
	
	From the bifurcation diagram shown in Figure \ref{fig3} (a), we observe that the $3\to 4$ transition occurs by a pitchfork bifurcation in the hyperbolic point at $x^*=0.5$. In the bifurcation, the hyperbolic point becomes elliptic and two new hyperbolic points emerge in the system. Before this isochronous transition, we have the scenario depicted in Figure \ref{fig3} (b), where we observe three islands identified with different color. The hyperbolic point at $x=0.5$ is highlighted by the presence of a separatrix structure. A post-transition scenario is shown in Figure \ref{fig3} (c), a phase space with four islands. We observe that the ``separatrix" structure disappears and a new island, centered at $x=0.5$, emerges in the phase space. The sequence of Figure \ref{fig3} represents the second type of possible transition, that is, a transition by pitchfork bifurcation with no intermediate mode.
	
\begin{figure}[!h]
		\begin{center}
			\includegraphics[width=1\textwidth]{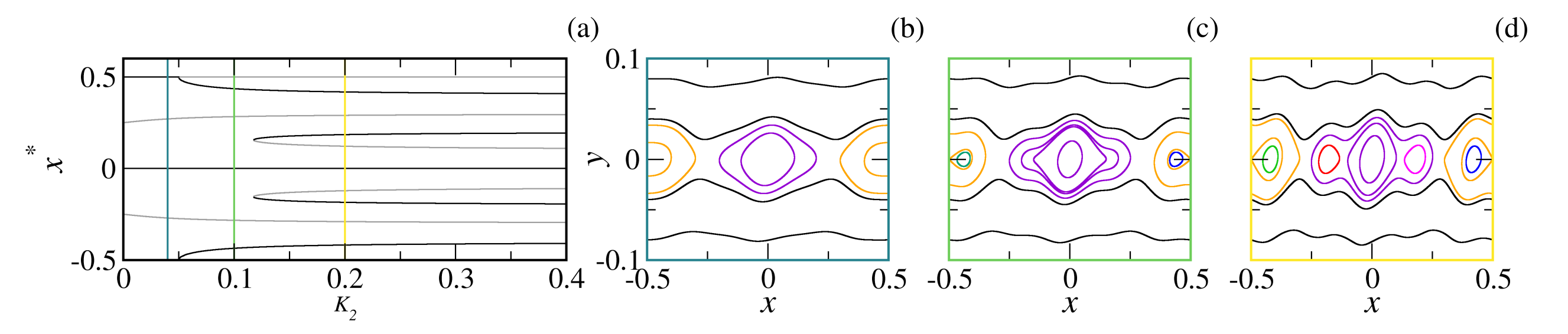}
			\caption{Transition with intermediate modes, for $m_1=2$, $m_2=5$ and $K_1=0.05$. The bifurcation diagram in panel (a) shows the evolution of the fixed points in the systems, where we observe first a pitchfork bifurcation followed by two saddle-node bifurcations. For the first mode, we have the phase space shown in  (b) with $K_2=0.04$. The intermediate mode is represented in (c) for $K_2=0.1$. Lastly, we have the predominance of the second mode in (d), where $K_2=0.2$.}
			\label{fig4}
		\end{center}
	\end{figure}

	Lastly, we investigate the $2 \to 5$ transition, which represent the third type of transition characterized by the presence of intermediate modes. The bifurcation diagrams and the respective phase spaces are shown in Figure \ref{fig4}.
	
	In the bifurcation diagram of Figure \ref{fig4} (a), we observe two isochronous bifurcations occurring at different values of $K_2$. First, there is a pitchfork bifurcation in $K_2=K_1=0.5$, where the elliptic point at $x=0.5$ becomes hyperbolic and two new elliptic points emerge in the system. As a result, an intermediate mode arises where three islands are present in the phase space. Secondly, we observe two saddle-node bifurcations at $K_2 \approx 0.13$, where two fixed points emerge around the elliptic point at $x=0$. These two saddle-node bifurcations complete the $2 \to 5$ transition.

	Just as in the case of Ref.\cite{mugnaine2024}, we analyze all the $m_1 \to m_2$ transitions with $m_1\in[1,5]$ and $m_2\in[m_1+1,6]$. All the bifurcation diagrams can be checked in the Supplementary Material \cite{sm}.

	\section{Conclusions}
	\label{section:conclusions}
	The emergence of isochronous islands occurs due to the superposition of distinct resonant modes interacting on the same winding number surface. In the two-harmonic standard map, the integers $m_1$ and $m_2$ identify the system's modes. Depending on the amplitudes $K_1$ and $K_2$, the system can exhibit between $m_1$ and $m_2$ islands, where $m_2 > m_1$.
	
	We observed the emergence of new islands as the
	amplitude of the second mode increases, forming three types of isochronous transitions. The first one occurs through saddle-node bifurcations, where $(m_2-m_1)$ pairs of elliptic and hyperbolic points emerge in the phase space for the same values of the parameters $K_1$ and $K_2$. The second route involves pitchfork bifurcations, where stable and unstable fixed points change their stability and two other fixed points with opposite stability emerge in the phase space. Lastly, we have the isochronous transition with intermediate mode, \textit{i.e.}, the $m_1\to m_2$ transition does not occur directly. For this last transition, we observe a combination of pitchfork and saddle-node bifurcations.
	
	\section{Acknowledgments}
	This research received the support of the Coordination for the Improvement of Higher Education Personnel (CAPES) under Grant No. 88887.320059/2019-00, 88881.143103/2017-01, the National Council for Scientific and Technological Development (CNPq - Grant No. 403120/2021-7, 311168/2020-5, 301019/2019-3), Fundação de Amparo à Pesquisa do Estado de São Paulo (FAPESP) under Grant No.  2024/03570-7 and 24/05700-5  and CNEN (Comissão Nacional de Energia Nuclear) under Grant No. 01341.001299/2021-54.
	
	\section*{Data availability}
	The source code and data are openly available online in the Oscillations Control Group Data Repository \cite{OCG}.
	

\end{document}